\documentstyle[12pt,epsfig]{article}
\def\lsim{\mathrel{\rlap{
\lower4pt\hbox{\hskip-3pt$\sim$}}
    \raise1pt\hbox{$<$}}}     
\def\gsim{\mathrel{\rlap{
\lower4pt\hbox{\hskip-3pt$\sim$}}
    \raise1pt\hbox{$>$}}}     

\begin{document}



\begin{center}
{\large {\bf SEARCH FOR A MIXED PHASE OF STRONGLY INTERACTING
MATTER AT THE JINR NUCLOTRON}}\footnote{
Contribution to the
Proceedings of the 8th International Workshop
"Relativistic Nuclear Physics: from hundreds MeV to TeV", JINR, Dubna,
May 23 - 28, 2005.}\\
\vspace*{5mm}

A.N.~Sissakian$^a$, \underline {A.S.~Sorin$^{a,}$}\footnote{
E--mail: {\tt sorin@theor.jinr.ru}}, M.K.~Suleymanov$^b$, V.D.~Toneev$^a$,
G.M.~Zinovjev$^c$ \\
\vspace*{3mm} {\it
$a)$ BLTP JINR, 141980 Dubna, Moscow region, Russia\\
$b)$ VBLHE JINR, 141980 Dubna, Moscow region, Russia\\
$c)$ BITP NAS, Kiev, Ukraine } \\
\end{center}

\vspace*{5mm}

{\small { \centerline{\bf Abstract}A physical program is
formulated for new facilities opening in Dubna for
acceleration of heavy ions with an energy up to  5 AGeV.}}\\[5mm]

\vspace*{3mm}

Over the last 25 years a lot of efforts have been made to search
for new states of strongly interacting matter under extreme
conditions of high temperature and/or baryon density, as predicted
by QCD. These states are relevant to understanding the evolution
of the early Universe after Big Bang, the formation of neutron
stars and the physics of heavy-ion collisions. The latter is of
great importance since it opens a way to reproduce these extreme
conditions in the Earth laboratory. This explains a permanent
trend of leading world research centers to construct new heavy ion
accelerators for even higher colliding energy.

Looking  at the list of heavy-ion accelerators, one can see that
after the first experiments at the Dubna Synchrophasotron,
heavy-ion physics successfully developed at Bevalac (Berkley) with
the bombarding energy to $E_{lab} \sim 2$ AGeV, AGS (Brookhaven)
$E_{lab} \sim 11$ AGeV, and SPS (CERN) $E_{lab} \sim 160$ AGeV.
The first two machines are closed now. The nuclear physical
programs at SPS as well as at SIS (GSI, Darmstadt, $E_{lab} \sim
1$ AGeV) are practically completed. The new relativistic heavy-ion
collider (RHIC, Brookhaven) is intensively working in the
ultrarelativistic energy range $\sqrt{s_{NN}}\sim 200$ GeV for
searching signals of the quark-gluon plasma formation. In this
respect, many hopes are related to the Large Hadron Collider (LHC,
CERN) which will start to work in  the TeV region in two-three
years. The low-energy scanning program at SPS (NA49 Collaboration)
revealed an interesting structure in the energy dependence of some
observables at $E_{lab} \sim 20-30$ AGeV, which can be associated
with the exit of an excited system into a decofinement state. This
fact essentially stimulated a new large project FAIR GSI
(Darmstadt) for studying compressed baryonic matter in a large
energy range of $E_{lab} =10-30$ AGeV which should come into
operation in 2013 year.

On the other hand, in JINR there is a modern superconducting
accelerator, Nuclotron, which has not realized its planned
parameters yet. In this talk  we would like to present arguments
that acceleration of  heavy ions like $Au$ at the Nuclotron up to
the maximal planned energy $E_{lab}=5$ AGeV, allows one to study
properties of hot and dense nuclear matter to be competitive at
the world level.

A convenient way to present a variety of possible states of
strongly interacting matter is a phase diagram in terms of
temperature $T$ and baryon chemical potential $\mu_B$ (or baryon
density $n_B$), as shown in Fig.1. From this schematic picture it
is seen in which region of the diagram the given phase is realized
and which colliding energies are needed to populate this region.
In particular, the boundary of the deconfinement $\&$ chiral
symmetry restoration  transition may be reached even below
energies planned in the FAIR GSI project but the nuclear matter
compression should be high enough.

This point is illustrated in more detail in Fig.2.

\begin{figure}
 \hspace*{3cm} \includegraphics[width=9cm]{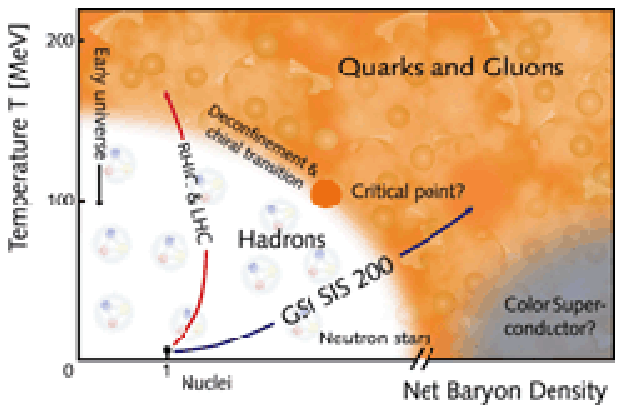}
  \caption{Artist's view of the phase diagram~\cite{GSI300} }
\end{figure}

\begin{figure}[h]
  \includegraphics[width=7.3cm]{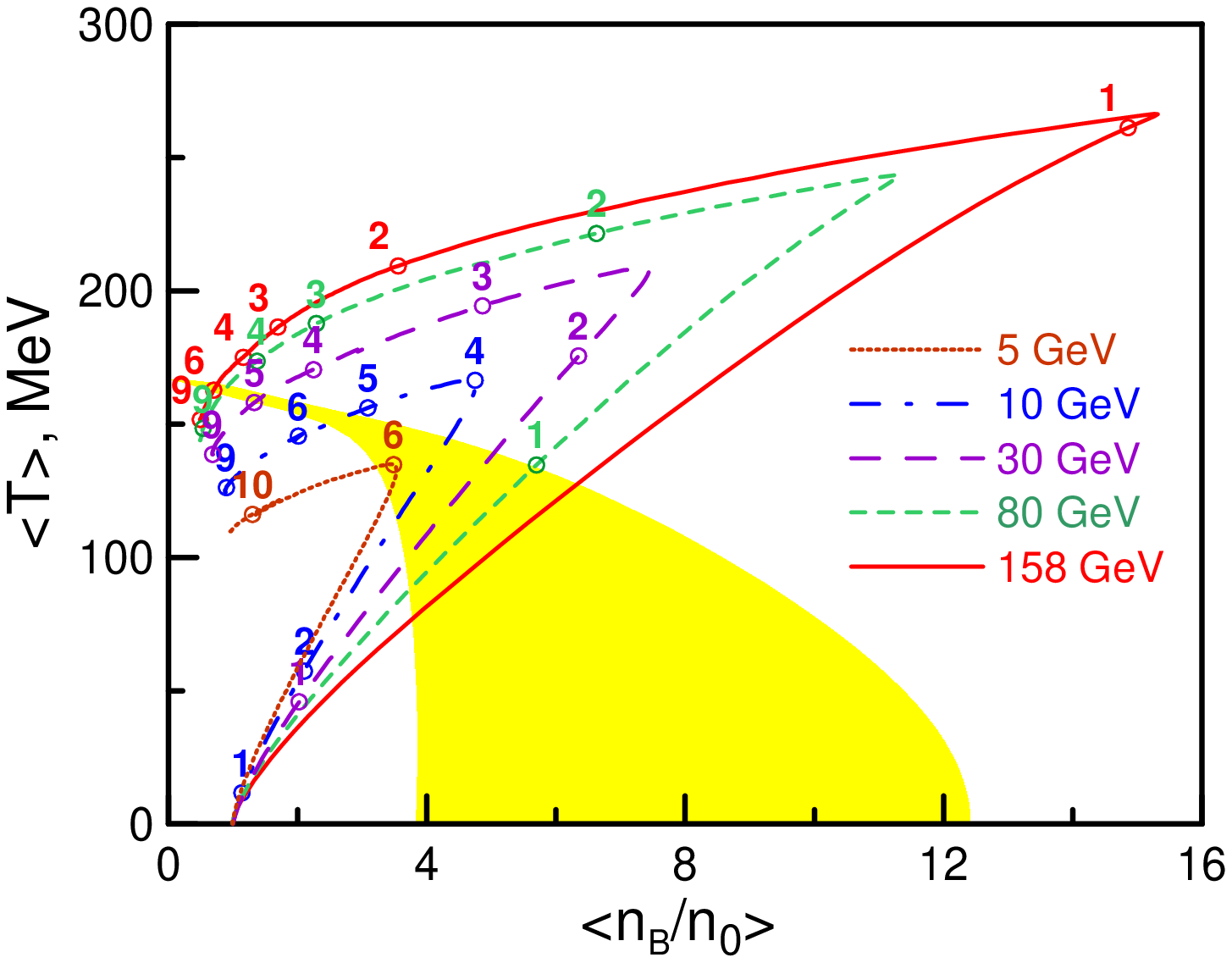}
  \hspace*{2mm}\includegraphics[width=7.3cm]{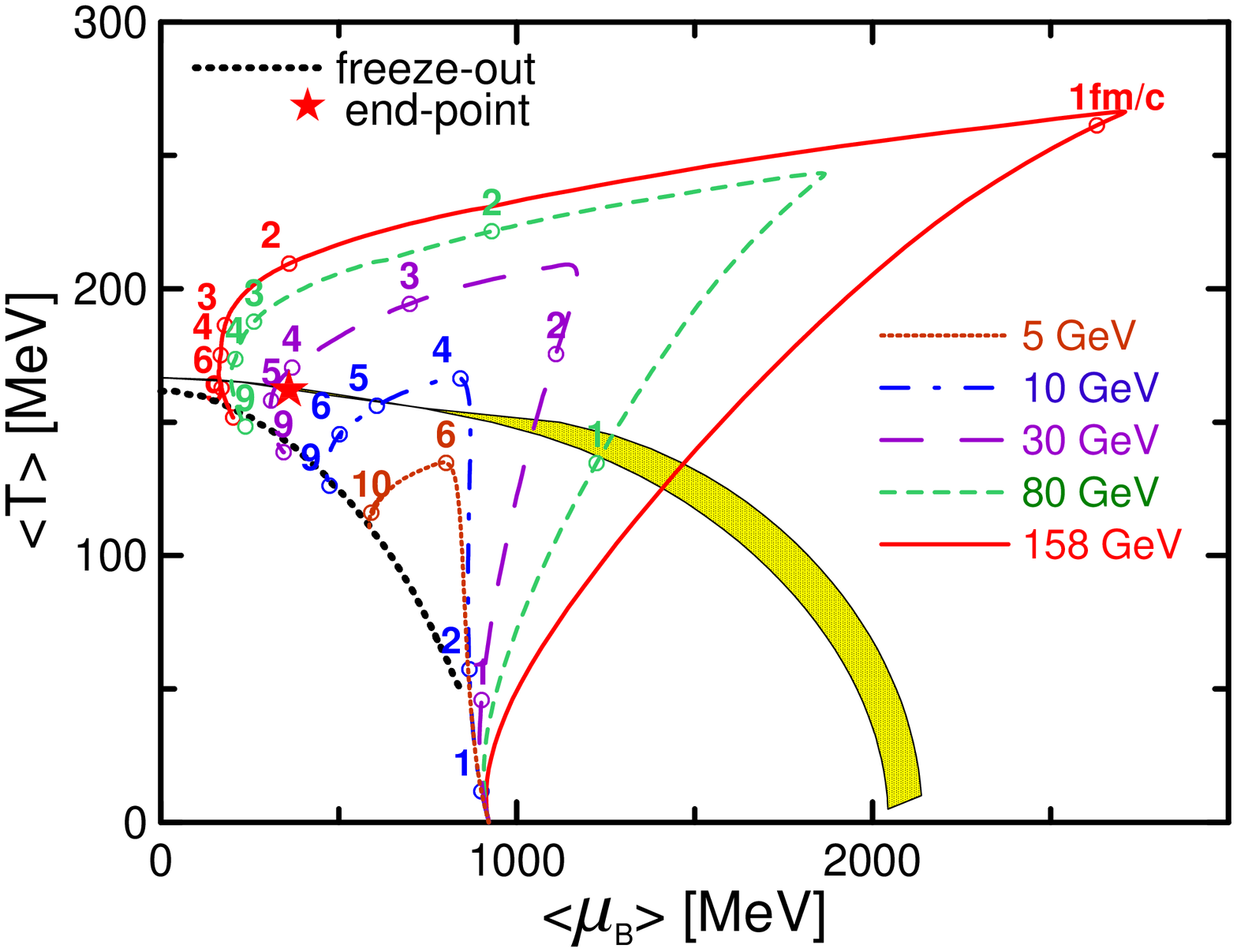}
  \caption{Dynamical trajectories for central $Au+Au$ collisions in $T-n_B$ (left panel)
  and  $T-\mu_B$ (right panel) plane for various bombarding energies calculated within
  the relativistic 3-fluid hydrodynamics~\cite{IRT05}. Numbers
  near the trajectories are the evolution time moment.  Phase boundaries are
  estimated in a two-phase bag model.
  }
\end{figure}

As is seen, a system, formed in a high energy collision, is fast
heated and compressed and then starts to expand slowly reaching
the freeze-out point which defines observable hadron quantities.
Indeed, at the maximal achievable Nuclotron energy $E_{lab}=5$
AGeV the system "looks" into the mixed phase  for a short time
(the left part of Fig.2). However, uncertainties of these
calculations are still large. If heavy masses for quarks are
assumed, the phase boundary is shifted towards higher $\mu_B$ (the
right part in Fig.2). On the other hand, dynamical trajectories
are calculated for a pure hadronic gas equation of state, and the
presence of a phase transition may noticeably change them. In
addition, near the phase transition the strongly interacting
system behaves like a liquid rather than a gas, as was clarified
recently at small $\mu_B$~\cite{ZS}. As to high $\mu_B$ values, it
is a completely open question. It is hard to believe that some
irregular structure like that at $E_{lab} \sim 30$ AGeV
~\cite{Gazd05} can manifest itself at the Nuclotron energy. So the
global observables (average multiplicities, rapidity spectra,
transverse distributions and so on) are expected to be quite
smooth with energy. However, it might not  be the case for more
delicate characteristics. In any case, due to the proximity of the
the phase diagram region under discussion to the confinement
transition and chiral symmetry restoration, some precursory
phenomena cannot be excluded at a bombarding energy about 5 AGeV,
which opens a new perspective for physical investigations at the
Dubna Nuclotron.

Below some arguments in favor of this statement are given:

\begin{itemize}
\item  Properties of hadrons  are expected to change in hot and/or
dense baryon matter~\cite{TK94,BR96}.  This change concerns
hadronic masses and widths, first of all for the $\sigma$-meson as
the  chiral partner of pions which characterizes a degree of
chiral symmetry violation and can serve as a "signal" of its
restoration as well as the mixed phase formation. Rare decays in
matter of vector mesons (particularly $\rho $ and $\omega$) are
also very attractive.
\begin{figure}[h]
 \hspace*{4cm} \includegraphics[width=8cm]{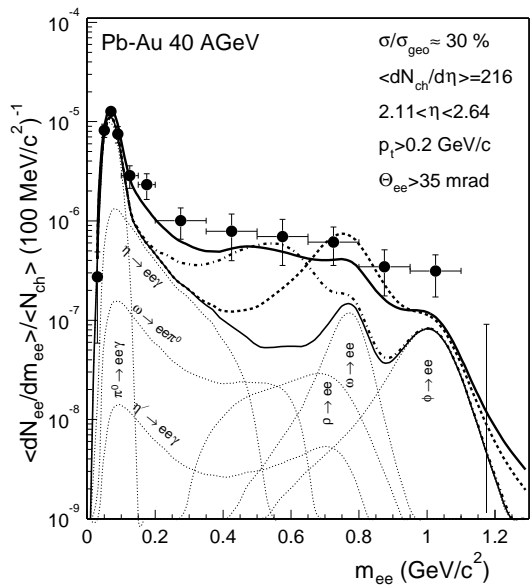}
  \caption{$e^+e^-$ invariant spectra from central $Pb+Au$ (40 GeV)
  collisions~\cite{CERES40}. Thin solid and dotted lines are hadronic
  cocktail and calculation results for free $\rho$ mesons, respectively.
  Appropriate thick solid and dash-dotted lines are calculated in the
  Rapp-Wambach~\cite{RW00} and Brown-Rho~\cite{BR96} scenarios.
  Contributions of different channels are shown, as well.}
\end{figure}

The presence of in-medium modification of $\rho$-mesons has been
proved in the CERES experiments, see Fig.3. The observed essential
enhancement of low mass ($0.2 \lsim M \lsim 0.7$) lepton pairs as
compared to free hadron decays, is due to the influence of hot and
dense nuclear matter on properties of the $\rho$-meson spectral
function. Unfortunately, poor resolution in the di-electron mass
does not allow discrimination of different physical scenarios of
this effect in the CERES experiments.

As to in-medium $\sigma$-meson decay, some indications were
obtained in reactions induced by pions and
$\gamma$'s~\cite{CHAOS,CBall,TAPS2}. In Fig.4, the relative pion
pair abundance $C_{\pi\pi}^A$, defined as
$$C_{\pi\pi}^A =\frac {
\sigma^A (M_{\pi\pi})}{ \sigma_T^A} \ / \ \frac {
\sigma^N(M_{\pi\pi})}{ \sigma_T^N}~,$$ is presented for different
isotopic states of observed pions versus the invariant pion pair
mass $M_{\pi\pi}$.
\begin{figure}[h]
 \hspace*{2cm} \includegraphics[width=12cm]{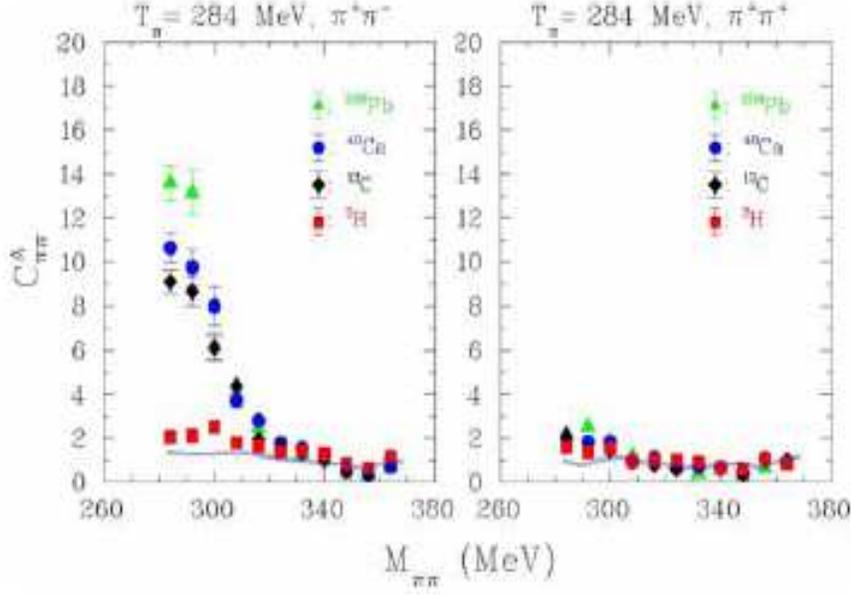}
  \caption{Invariant mass distribution of pion pairs from $\pi$-nucleus
  interactions~\cite{CHAOS}. }
\end{figure}
A sizable enhancement is observed at low $M_{\pi\pi}$ in the case
of $\pi^+ \to \pi^-\pi^-$ reaction but not for the $\pi^+ \to
\pi^+\pi^+$ case~\cite{CHAOS}. This enhancement is related to a
possible $\pi\pi$ scattering in a nucleus via formation of a
virtual scalar $\sigma$ meson which is forbidden for the
$\pi^+\pi^+$ state. The effect is getting stronger for heavier
nuclei. Recently, using ELSA tagged facilities in Bonn the
in-medium modification of the $\omega$-meson has first been
observed in reaction $\gamma +A \to \omega +X \to \pi^0+\gamma
+X$~\cite{TAPS2}. A small shift in the $\omega$ mass was detected.
Note, however, that in $\pi$- and $\gamma$-induced reactions we
deal with comparatively low baryon density states, $n_ B\sim
(1-2)n_0$. There are no similar experiments with heavy ions.

Nevertheless, there are theoretical proposals to probe chiral
symmetry restoration in the vicinity of the phase transition
boundary. In particular, it was shown~\cite{CH98,VKBRS98} that a
two-photon decay of the $\sigma$-meson formed as an intermediate
state in $\pi\pi$ scattering may be a very attractive signal. As
pictured in Fig.5, at temperature in the vicinity of
\begin{figure}[h]
 \hspace*{3.cm} \includegraphics[width=7.5cm,angle=-90]{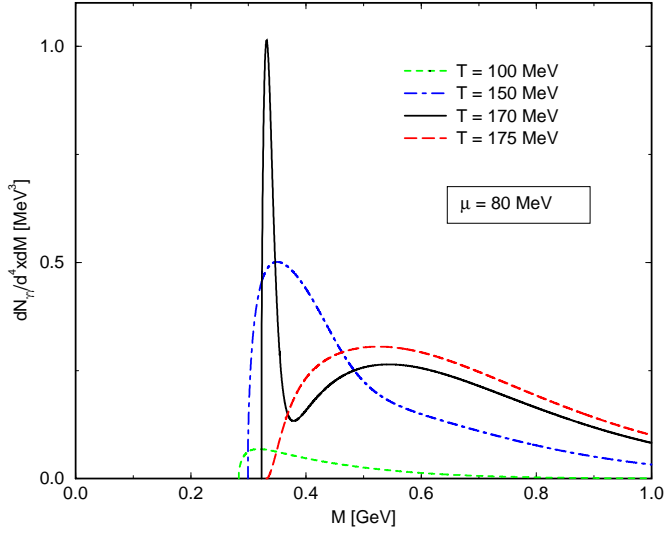}
  \caption{Invariant mass spectra of $2\gamma$ at $\mu_B=80$ MeV
  and different temperatures~\cite{VKBRS98}. }
\end{figure}
 the phase transition, when $m_\sigma \sim 2m_\pi$, there is an anomalous peak
in invariant mass spectra of $\gamma$ pairs which may serve as a
signal of the phase transition and formation of a mixed phase.
Certainly, there is a huge combinatorial background due to $\pi^0
\to \gamma\gamma$ decays, but the Nuclotron energy is expected to
have some advantage against higher energy accelerators because the
contribution of deconfined quarks-gluons will be negligible.

This effect may be observed in the $e^+e^-$ decay channel, as
well~\cite{Weld92}.

 \item Electromagnetic probes discussed above  carry out
 information concerning the whole evolution of colliding nuclei and states which
 are realized at that. The bulk of produced hadrons is related to
 the freeze-out point where information on the interaction dynamics
 has been essentially erased. So the expected behavior of global
 hadronic observables is smooth. However, some peculiarities of delicate
 hadron characteristics may be found and their hints are available
 even now.

\begin{figure}[h]
 \hspace*{4cm} \includegraphics[width=9cm]{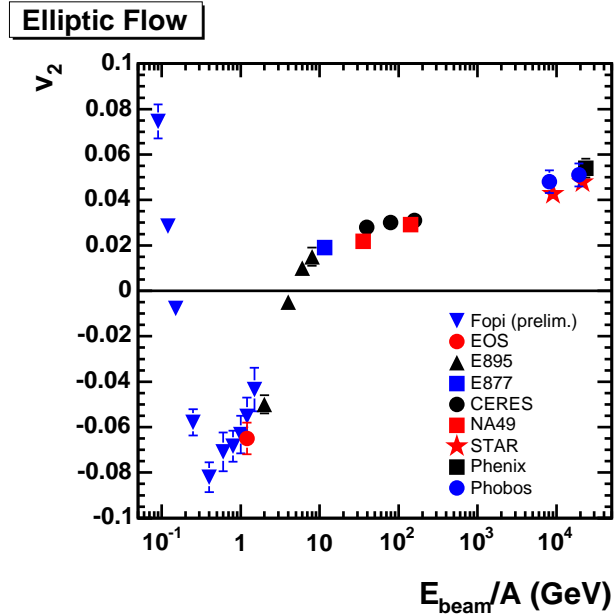}
  \caption{Excitation function of the proton elliptic flow ~\cite{St04}. }
\end{figure}

 In Fig.6, the beam energy dependence of the elliptic flow coefficient $v_2$
  is presented for protons in the midrapidity  range from noncentral
  heavy-ion collisions~\cite{St04}. The elliptic
 flow characterizes the angular anisotropy  in the transverse
 momentum plane as
 $$ dN/d\phi \sim \left[ 1+2v_1 \cos (\phi)+2v_2 \cos (2\phi) \right]~, $$
 where the $\phi$ angle is measured from the reaction plane. As is
 seen,  just near the  Nuclotron energy $E_{lab}\sim 5$ AGeV the
 coefficient $v_2$ changes its sign and the transverse flow
  evolves from the out-of-plane towards in-plane emission. This fact may be treated
  as softening of the equation of state to be a precursor of the phase transition.
 The elliptic flow shows an essentially linear dependence on the
impact parameter with a negative slope  at $E_{lab}=2$ AGeV, with
a positive slope at 6 AGeV and with a near zero slope at 4
AGeV~\cite{Ch01}. This dependence serves as an important
constraint to  high-density behavior of nuclear matter for
discrimination of various equations of state.

\item Strangeness enhancement is an intriguing point of physics of
heavy ion collisions, being one of the first proposed signals of
quark-gluon plasma formation. An important experimental finding is
the observation of some structure ("horn") in the energy
dependence of reduced strangeness multiplicity at $E_{lab}\sim 30$
AGeV, predicted in~\cite{GG99} as a signal that the formed excited
system came into a deconfinement phase. As an example, in Fig.7
the $K^+/\pi^+$ average multiplicity ratio is displayed as a
function of the bombarding energy. The "horn" structure is well
visible and is getting even more prominent if the recent
measurement of $<K^+>/<\pi^+>\simeq 0.22$ at $E_{lab}=20$
AGeV~\cite{Gazd05} is
\begin{figure}[h]
 \hspace*{4cm}\includegraphics[width=8cm]{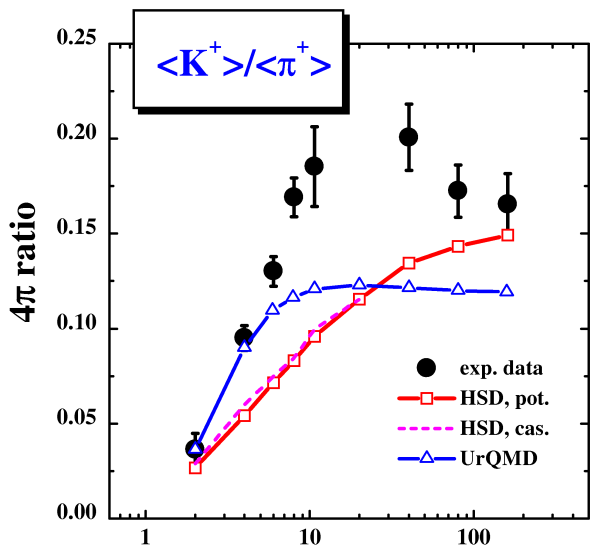}
   \caption{Energy dependence of the relative strangeness abundance of $K^+$
    mesons~\cite{WBCS02}. Curves are results of different transport calculations. }
\end{figure}
additionally plotted and the bombarding energy is presented in the
logarithmic scale (which is quite natural when RHIC data are
supplemented). It is of great interest that these global
characteristics are not explained by modern transport theory
(UrQMD, HSD models). While the average pion and kaon
multiplicities are well reproduced at the SIS and SPS energies,
the above-mentioned models essentially underestimate the $K /\pi$
ratio in the AGS energy domain. It is remarkable that the
divergence between the transport theory and experiment starts just
at the Nuclotron energy. This increases interest in future
experiments at the Nuclotron.

\item One should note that peculiarities at the same "horn" energy
are observed also in other measured quantities~\cite{Gazd05}. The
other example is given in Fig.8 where an average pion number per
nucleon-nucleon interaction is plotted.

\begin{figure}[h]
 \hspace*{4cm}\includegraphics[width=10cm]{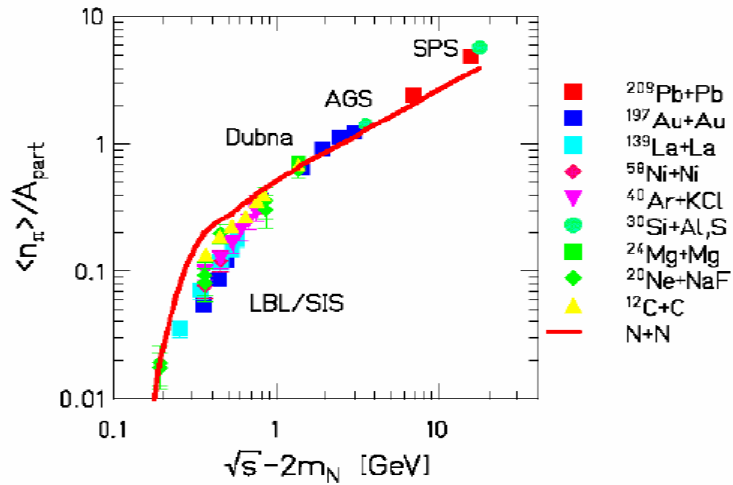}
   \caption{Pion multiplicity per participating nucleon for
   nucleus-nucleus (symbol) and nucleon-nucleon collisions (solid
   line) as a function of available energy in nucleon-nucleon collisions~\cite{SS99}.
    }
\end{figure}

 The reduced pion multiplicity $<n_\pi> / A_{part}$ for
 nucleus-nucleus collisions is below  that for elementary $NN$ interactions at moderate
 energies and exceeds approprate $NN$ values in a relativistic
 domain. According to~\cite{Gazd05,GG99}, the equality of the nuclear and elementary
 $<n_\pi> / A_{part}$ ratios at the "horn" energy is treated as an
 argument in favor of the onset of a deconfinement. However, this
 proximity of nuclear and elementary reduced pion multiplicities
 starts at the Dubna Nuclotron energy and could naturally be explained by the
 role of the $\Delta$ isobar in the pion absorbtion. To clarify this or
 alternative interaction mechanisms, new experimental data for heavy ion
collisions with scanning over the Nuclotron energy range are
needed.

\item   Existing experimental data indicate a large degree of
equilibration of nuclear matter in high-energy heavy-ion
collisions~(see review-article \cite{BRS03}).
 The mechanisms of equilibration are still unclear. In particular, a large difference between dynamical and
 statistical models is observed for heavy baryons (e.g. for multi-strange hyperons, $\Xi$  and $\Omega^-$
 and resonances) measured at high energies. The role of multiparticle interactions is also obscure.
 The corresponding data for the Nuclotron energies are extremely poor. Future Nuclotron measurements
 could give a strong input for a further development of
 dynamical and statistical approaches and allow one to disentangle  these two types of models.

To a certain extent, the  study of fluctuations in relativistic
strongly interacting matter may help in solving the equilibration
problem. Experimental data on event-by-event fluctuations (e.g.
 fluctuations of particle multiplicity, electric, baryon and strangeness charges)
  in nuclear collisions give a unique
possibility to test recent ideas on the origin of fluctuations in
relativistic interacting systems~\cite{Fluct}. Measurements
require tracking detectors of  large
 acceptance and precise control of collision
centrality on event-by-event basis. Up to now only results on very
limited acceptance at high energies are available, thus  new
measurements at the Nuclotron energy are of particular importance.
From the experimental point of view the Nuclotron energy range
seems to be ideal for these measurements. This is because moderate
particle multiplicity and their relatively broad angular
distribution simplify an efficient detection of all produced
charged particles.

An investigation of the narrow Hanbury-Brown-Twiss
correlations~\cite{HBT} joins this problem. The HBT analysis
exploits particle correlations at small relative velocities and is
widely used to study space-time characteristics (the system size,
lifetime, freeze-out duration, expansion time) of the production
processes in relativistic heavy-ion collisions.  There is
substantial experience accumulated at JINR in the field of
particle correlations. Such HBT data
 are missing at $E_{lab}<15$ AGeV.

One should emphasize that all these investigations suppose that
centrality of heavy-ion collisions is under control and centrality
scanning of the characteristics under discussion is an
indispensable  condition.

Measurements of these quantities at the Nuclotron energies should
be considered as a necessary continuation of global efforts to
establish the energy dependence of properties of hadron production
and search for  signals of a phase change in nuclear collisions.

\end{itemize}

In this regard the following theoretical and experimental studies
at JINR are considered as perspective:
\begin{enumerate}
\itemsep=-2mm \item[1)] research into the hadron properties in hot
and/or dense baryonic matter. A spectral function change is
expected, first of all for the $\sigma$-meson as a chiral partner
of pions, which characterizes a degree of chiral symmetry
violation. The rare specific channels of $\rho$-meson decays are
also quite attractive.
\end{enumerate}

\vspace*{-.3cm} \noindent
 Solving these issues assumes a proper
understanding of reaction mechanisms of high-energy colliding
ions, knowledge of properties of interacting matter and its
equation of state. In this respect more general researches are in
order:

\vspace*{-.1cm}

\begin{enumerate}
\itemsep=-2mm \item[2)] analyzing multiparticle hadron
interactions, targeted to the development of new statistical
treatment as well as codes for space-time evolution of heavy
nuclei collisions at high energies. Particular attention should be
paid to signals of new phase formation during this evolution;

\item[3)]  studying the system size, lifetime, freeze-out
duration, expansion time in the HBT analysis (noticeable volume
expansion is expected if the mixed phase is formed), scanning in
atomic number and energy;

\item[4)]  analyzing the energy and centrality dependencies of the
pion, hadron resonance and strange particle multiplicities, and
the ratio of their yields, together with the transverse momentum,
including $K^-, K^*-$ and $\phi$-meson spectra as well as
manifestation of baryon repulsion effects on hadron abundances;

\item[5)] studying di-leptons (electron and muon pairs) production
to see in-medium modification of hadron properties at high baryon
densities;

\item[6)] studying angular correlations in the transverse plane
as well as radial, directed and elliptic flows;

\item[7)]  analyzing fluctuations of multiplicities, electric
charge and transverse momenta for secondary particles (their
energy dependencies could give information on the phase transition
range);

\item[8)] analyzing nuclear fragments characteristics versus
centrality (change of behaviour comparing to the peripheral
collisions is expected), universality of nuclear fragmentation;

\item[9)]  energy and atomic number scanning for all
characteristics of central heavy nuclei collisions (this might
allow one to obtain information on the equation of state of
strongly interacting matter in the transition area), difference
between central collisions of light nuclei and peripheral heavy
ion collisions.
\end{enumerate}

The JINR Nuclotron has a possibility to accelerate heavy ions (up
to $A>$200) to the maximal energy of 5 AGeV in about a year. This
gives a chance to address experimentally many recent problems
within the next several years before the FAIR GSI accelerator
come into operation. The proposed research program at the
Nuclotron may be considered as a pilot study preparing for
subsequent detailed investigations at SIS-100/300~\cite{GSI300}
and as an integral part of the world scientific cooperation to
study the energy dependence of hadron production properties in
nuclear collisions.

\vspace*{3mm}
 {\bf Acknowledgements}
 We would like to thank members of the working group N. Amelin,
K. Abramyan, D. Blaschke, Yu. Bystritsky, Yu. Kalinovsky, V.
Karnaukhov, A. Kovalenko, V. Krasnov, E. Kuraev, A. Kurepin, R.
Lednicky, J. Lukstins, A. Malakhov, I. Molodtsova, S. Molodtsov,
A. Radzhabov, V. Skokov, O. Teryaev, A. Vodopianov, M. Volkov and
V. Yudichev,  for a fruitful collaboration. We greatly appreciate
many useful and valuable discussions with M. Gazdzicki, M.
Gorenstein, H. Gutbrod T. Hatsuda, T. Kunihiro, H. Satz and H. Str\"obele.
We would like also to express special thanks to A. Filippov, V.
Voronov and V. Zhuravlev for support during these investigations.
This work was supported in part by RFBR Grant N 05-02-17695.

\end{document}